# Photon-number resolving detector based on a series array of superconducting nanowires

S. Jahanmirinejad,[1,a)] G. Frucci,[1] F. Mattioli,[2] D. Sahin,[1] A. Gaggero,[2] R. Leoni,[2] and A. Fiore[1]

[1]COBRA Research Institute, Eindhoven University of Technology, PO Box 513, 5600 MB Eindhoven, The Netherlands

[2]Istituto di Fotonica e Nanotecnologie, CNR, Via Cineto Romano 42, 00156 Roma, Italy

[a)] Electronic mail: s.jahanmiri.nejad@tue.nl

We present the experimental demonstration of a superconducting photon number resolving detector. It is based on the series connection of $N$ superconducting nanowires, each connected in parallel to an integrated resistor. The device provides a single voltage readout, proportional to the number of photons absorbed in distinct nanowires. Clearly separated output levels corresponding to the detection of $n$=1-4 photons are observed in a 4-element detector fabricated from an NbN film on GaAs substrate, with a single-photon system quantum efficiency of 2.6% at $\lambda$=1.3µm. The series-nanowire structure is promising in view of its scalability to large photon numbers and high efficiencies.

Conventional optical detectors generate an electrical signal proportional to the intensity of the incident light. However their sensitivity is limited by the electrical noise in the amplification circuit. On the other hand, single photon detectors (SPDs), which are extremely sensitive devices, usually show a strongly nonlinear response, i.e. their output signal level is independent of the number of photons that simultaneously hit the detector. The gap between these two detection regimes can be filled with a photon-number-resolving (PNR) detector, a device as sensitive as an SPD, and with a capability of precisely determining the number of photons that strike the detector in a very short time window (smaller than the detector response time). Realization of PNR detectors would benefit many applications from linear optical quantum computing[1] to near-infrared spectroscopy.

In the last years, several detector technologies based on combining linear operation and single-photon sensitivity in one device have been demonstrated. However, in the telecommunication wavelength region, which is interesting for many applications, their performance has remained limited in terms of speed and dynamic range. For instance, transition-edge-sensors operated at sub-Kelvin temperatures are plagued by slow response times[2], while self-differencing InGaAs avalanche photodiodes (APDs) offer limited photon-number discrimination ability[3]. In another approach, PNR detection can be achieved by combining multiplexing techniques and SPDs. Time-multiplexed APDs have been reported, but with a necessarily reduced count rate due to the delay loops[4,5]. Frequency up-conversion combined with Silicon photomultiplier (SiPM) provides large dynamic PNR but with added noise due to up-conversion and cross-talk between the pixels, which requires detector calibration[6]. An array of nanowire superconducting single photon detectors (SSPDs) has been proposed to operate as a PNR detector based on spatial multiplexing, using either a separate readout for each element[7] or a parallel configuration with single output[8]. These implementations take advantage of the very high sensitivity, short dead time, and low timing jitter of superconducting nanowire detectors in the near-infrared range[9]. A detector geometry allowing a single readout of the photon number is particularly promising in view of simplicity of use and scalability to large photon numbers. However, the parallel nanowire detector (PND) explored so far is limited in terms of dynamic range (maximum number of photons which can be detected in a pulse), and efficiency due to the problem of current redistribution in the array after photon detection in one or more wires. The current redistribution issue can result in spurious switching of the wires which did not absorb a photon, generating false counts[10]. Avoiding this issue requires decreasing the number of elements in the array as well as limiting the bias current well below the critical current, which prevents reaching the highest efficiency. Recently, we have proposed[11] a device structure called series nanowire detector (SND), which is the electrical dual of PND and is designed to solve the current redistribution problem. It is based on the series connection of $N$ nanowires, each connected in parallel to a resistor ($R_P$) as shown in Fig. 1(a). All the detecting sections are equally biased (in series) with a bias current ($I_B$) close to the critical current ($I_C$). Upon absorption of a photon in one section (firing), a large resistance develops in the wire and the current is diverted into the parallel resistor producing a voltage pulse. If more photons are absorbed in distinct sections, the voltages produced across them are summed up at the output, resulting in a voltage proportional to the number of incoming photons. Firing of a wire in the SND *reduces* the bias current in the other unfiring wires, as compared to the PND where the current in the unfiring wires is increased. This prevents the uncontrolled switching of the unfiring wires as it may happen in the PND if the bias current is set very close to the critical current. Therefore using the series configuration can result in higher QE and the possibility to scale to large photon numbers[11]. The device performance in terms of output amplitude and speed can be further optimized by means of a high-impedance preamplifier which is easily realized using a high electron mobility transistor mounted next to the SND and operated cryogenically. In order to illustrate the operation principle of SND's photoresponse, Fig. 1(b) shows the simulated output voltage of a 4-element SND when $n$=1-4 photons are detected in distinct elements. The simulation parameters are based on the model described in Ref. 11 and on the device parameters used in the experiment described below ($I_C$=12.5μA, $I_B$=0.99$I_C$, $R_P$=29.5Ω, $R_L$=50Ω, kinetic inductance $L_K$=100nH for each element). The relevant recovery time constant is approximately given by $\tau_F = NL_K/(NR_P \| R_L)$ [11] and is calculated as $\tau_F$=11.4ns in this device.

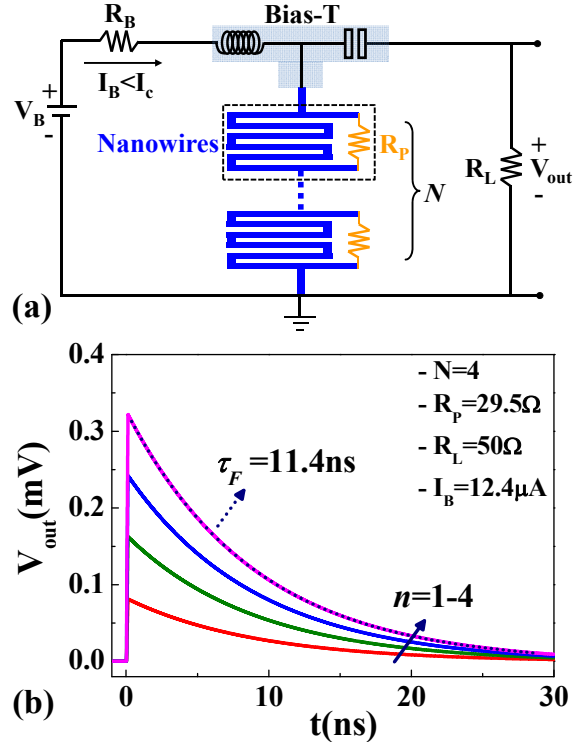

FIG. 1. (a) Schematics of the electrical circuit and the layout of an *N*-element SND structure. (b) Transient response of a 4-SND biased at $I_B$=0.99$I_C$, simulated using the electrothermal model[11] with $R_P$=29.5Ω and $R_L$=50Ω.

In this work, the fabrication and experimental demonstration of an SND with four detecting elements in series (4-SND) is reported, as a proof of principle. SNDs are fabricated from a 4.5nm thick NbN film grown on GaAs substrate by reactive magnetron sputtering, at relatively low deposition temperature[12]. This particular film exhibited a superconducting transition temperature of 9.5K and a superconducting transition width of 0.7K. To fabricate the SND structures, four nanolithography steps are carried out by using field emission gun electron beam lithography (EBL) system with acceleration voltage of 100kV. In the first step, Ti/Au (60nm Au on 10nm Ti) contact pads together with alignment markers are fabricated by lift-off using a Polymethyl

Methacrylate (PMMA) stencil mask. This is followed by the second lithographic step to define the thin Ti/Au (20nm Au on 5nm Ti) pads used for the electrical contact of the resistors. In the third step, hydrogen silsesquioxane (HSQ) is used as an etch mask to pattern the meanders with reactive-ion-etching (RIE). In the last step the resistors (40nm thick Au/Pd film) in parallel with the nanowires are fabricated by lift-off via a PMMA stencil mask. Fig. 2 shows a scanning electron microscope (SEM) image of the fabricated SND with $N$=4 detecting elements. The NbN nanowires are 100nm wide, covering a total active area of $12\times12\mu m^2$ with a filling factor $f$=40%.

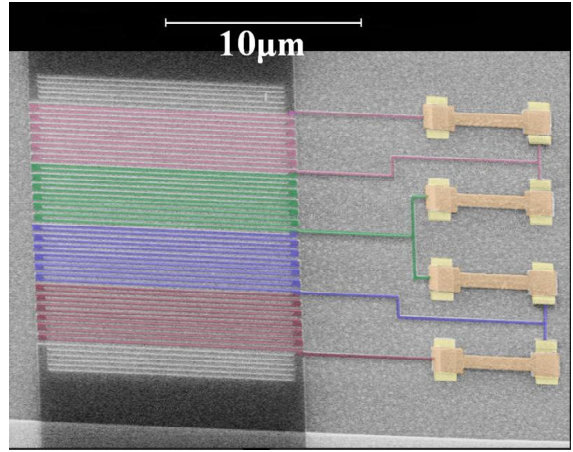

FIG. 2. Scanning electron microscope (SEM) image of the SND with $N$=4 detecting elements (4-SND) in series fabricated from a 4.5nm thick NbN film on GaAs substrate. The 100nm NbN nanowires cover the total active area of $12\times12\mu m^2$ with $f$=40% fill factor. The active wires, the parallel resistors and their contact pads have been colored for clarity.

We performed the electro-optical characterization of the SND in a closed-cycle cryocooler with a base temperature of 1.18K on the experimental plate and stability within 0.01K. As it is schematically depicted in Fig. 1, the bias current is supplied through the DC port of a bias-T by a voltage source in series with a small bias resistor. For the electrical characterization, the bias-T is solely used as a lowpass filter to remove the noise of the bias source, while its RF arm is terminated with a 50Ω matched load. The I-V characteristic of the device is displayed in Fig. 3.

In the superconducting state, the parallel resistors are short circuited by the zero resistance of the wire, having no effect in the I-V. As soon as the bias current exceeds the critical current of a nanowire, a normal domain with finite resistance is formed across it, and a part of the bias current is pushed to the parallel resistance. Due to the effect of the parallel resistance, the typical relaxation oscillation regime and the hotspot plateau are not observed. In a section, when the entire length of a nanowire becomes normal, the resistance is the parallel equivalent of the nanowire normal resistances ($R_{nanowire}$~70kΩ) and $R_P$, which is very close to $R_P$. Following the transition of all the sections to the normal state, the device has a resistance equal to the sum of all parallel resistances. The reciprocal of the slope of the linear fit indicated in Fig. 3 corresponds to $4R_P$=118Ω, hence $R_P$ is estimated to be 29.5Ω.

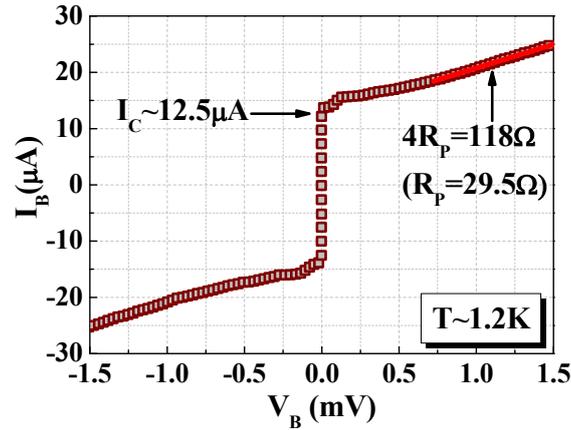

FIG. 3. IV characteristics of the 4-SND. The critical current is $I_C$~12.5µA at T=1.2K and the parallel resistance of each element is $R_P$~29.5Ω estimated from the slope of the fitted linear line.

For the optical characterization, the light produced by a pulsed laser-diode (50ps pulse width, 1.3µm wavelength, 10MHz repetition rate) was fed to the device through a single-mode polarization maintaining lensed fiber ($1/e^2$ beam diameter of 5µm) mounted on XYZ piezo stages providing an active alignment of the fiber on the detector area. The electrical response of the detector was collected through the RF arm of the bias-T, amplified using a chain of 50Ω, room-temperature amplifiers and directed either to a 40-GHz bandwidth sampling oscilloscope, or a 350-MHz counter for optical characterization.

In order to investigate the photon number resolving capability of the detector, the distribution of the output voltage pulses was measured using a sampling oscilloscope triggered with the laser-diode modulation signal. In order to couple the light equally to all sections of the SND, the beam was defocused to a larger spot diameter (estimated to be ~20μm), achieving uniform illumination across the 12x12μm$^2$ area of the detector. Fig. 4(a) shows as an example an oscilloscope trace at $I_B$=8.3μA and an average incident flux of ~3800 photons/pulse (the probability of single-photon detection in each element in this situation is ~0.076%). Discrete voltage heights corresponding to the detection of $n$=1-4 photons can be observed. The symmetric pulse shape (especially the slow rise time) as compared to the calculated response in Fig. 1(b), is ascribed to the effect of a low-pass filter (DC-80MHz) used to remove the high frequency noise of the signal. The inset in Fig. 4(a) shows an oscilloscope trace taken without using the lowpass filter for a higher photon flux so that only the 4-photon detection is observed. The simulated output voltage when all of the four detection sections of SND fire is also plotted (continuous line), showing good agreement with the experimental data. The minor discrepancies in the rise/fall times probably originate from the effect of the amplifier bandwidth which is not taken into account in the simulation. The related histogram for the recorded peak voltage levels is shown in Fig. 4(b), featuring a series of peaks broadened due to the presence of electrical noise. The histogram can be fitted with a multiple-peak Gaussian distribution (solid light blue line) where each peak, corresponding to the photon absorption in different sections of the SND, has a Gaussian distribution (gray dotted curves). The first peak which is entirely due to the electrical noise corresponds to the case when no photon was detected. The peaks at 12.5, 24.5, 37, and 50mV correspond to the detection of 1, 2, 3, and 4 photons, respectively. The evenly-spaced, clearly resolved peaks in the histogram show the photon-number-resolving functionality of the SND structure and confirm that conventional 50Ω amplification is sufficient for the 4-SND to distinguish between 1-4 photons present in an optical pulse. The observed increase in the width of the Gaussian fits with increasing photon number is likely due to the variation in the electrical noise coming from the varying device impedance in different detection events.

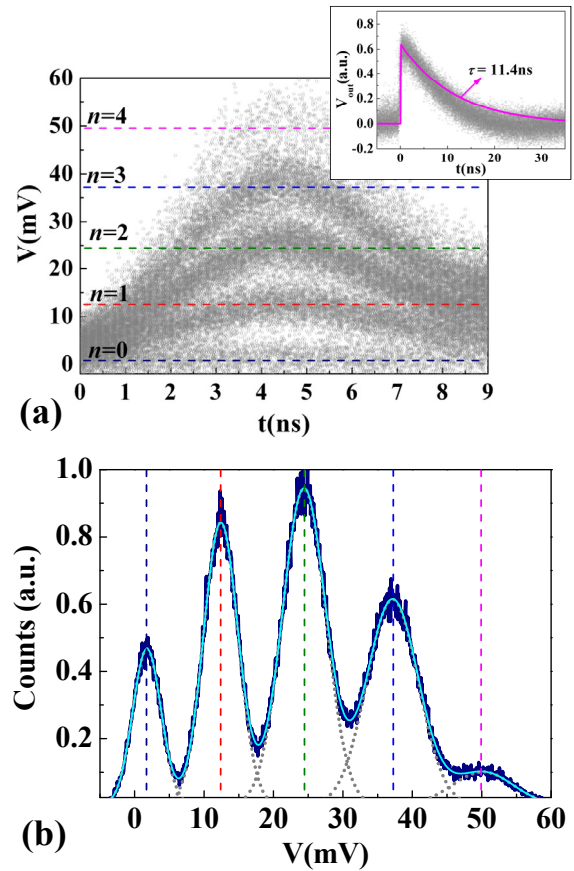

FIG. 4. (a) Photoresponse transients taken with a 40 GHz sampling oscilloscope showing four discrete amplitudes related to four detecting elements of the SND biased with $I_B$=8.3μA. Inset: Oscilloscope trace at a high input light power related to detection of 4 photons in each pulse (without using the lowpass filter), together with the simulated output voltage (n=4) (b) Measured (dark blue) and fitted (light blue) probability distribution of the detector output pulses corresponding to 0,1,2,3, and 4 photon detection events together with the Gaussians fit (gray dotted line) related to each peak.

In order to confirm that the four observed peaks are related to the detection of 1-4 photons, the photocount statistics were measured using a 350MHz counter. The photoresponse of the SND was sent to the counter and the photocounts were measured with the counter threshold level set to the midpoint between two of the detection levels observed in Fig. 4, for many different light powers. In this configuration, a count is measured when $n \geq n_{th}$ photons are detected. For illumination with a weak pulsed laser source such that the average number of detected photons $\mu_{det} \ll 1$, the count rate is expected to scale as $R(n_{th}, \mu_{inc}) \propto \mu_{det}^{n_{th}} = \eta^{n_{th}} \mu_{inc}^{n_{th}}$ where $\eta$ is the quantum efficiency and $\mu_{inc}$ is the average number of incident photons. In Fig. 5 the measured count rates relative to one-, two-, three-, and four-photon absorption events are plotted as a function of $\mu_{inc}$ in log-log scale. The solid lines are the linear fits to the data with slopes very close to 1, 2, 3, and 4, which confirms that the detector responds to 1-4 photon absorption events.

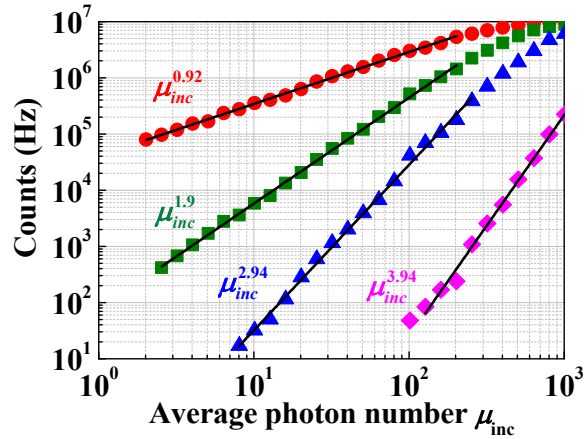

FIG. 5. Count rates measured at $I_B$=10μA relative to the ≥1 (red circles), ≥2 (green squares), ≥3 (blue triangles), and ≥4 (pink diamonds) photon detection events as a function of the mean incident photon number per pulse $\mu_{inc}$.

To characterize the SND in terms of efficiency, the single-photon system quantum efficiency (SQE) was measured under weak, well-focused illumination such that only 1-photon detection event was observed. The value of SQE was obtained by dividing the number of photocounts (corrected for dark counts) by the average number of photons at the input of the cryostat. The result is plotted in Fig. 6 (left axis, red spheres) as a function of bias current. The SQE attains its highest value of 2.6% at a bias current of $I_B=12.4\mu A$ ($0.99I_C$) with the light polarized parallel to the wires. Considering the measured fiber coupling efficiency of 95% and the calculated absorptance of 9% in a 4.5nm thick NbN grating (with 40% filling factor) on GaAs substrate[12], only 8.5% of the input light can be absorbed in the wires with top illumination at $\lambda=1.3\mu m$. Therefore, an intrinsic quantum efficiency (the ratio of detected to absorbed photons) of 30% is derived for the device. The efficiency can be enhanced both by improving the superconducting properties of the thin film and also applying advanced optical structures, such as optical cavities or waveguides[12-16].

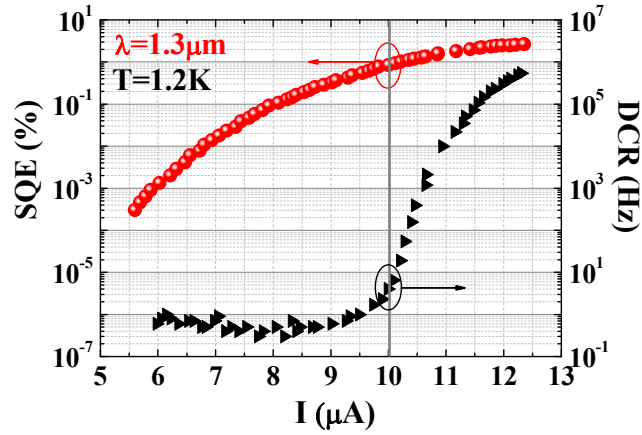

FIG. 6. Single-photon system quantum efficiency of an SND at $\lambda=1.3\mu m$ (left axis, red spheres) and dark count rate (right axis, black triangles), as a function of the bias current, measured at T=1.2K. The maximum SQE is 2.6% at 12.4μA. The solid vertical line shows a bias current at which SQE of 1% and DCR of 5Hz is obtained.

On the right axis of Fig. 6, the dark count rate (DCR) is presented as a function of the bias current (black triangles), as measured by blocking the optical input to the cryostat and moving the fiber away from the device to suppress any spurious light input. A low background level of 1Hz, due to electrical noise is measured at low bias currents, with a sharp increase for $I_B>10\mu A$.

At $I_B$=10μA, a SQE of 1% and DCR of ~5Hz are obtained. The fact that the device could be operated at 99% of the $I_C$ indicates the correct functionality of the series wire configuration, as predicted by our model[11] and therefore the possibility of reaching very high QEs. Furthermore, the timing jitter of the detector, after de-convolving the jitter from the laser, was measured to be 80ps.

In conclusion, we have successfully implemented a design for PNR detectors based on the series connection of *N* superconducting nanowires, each shunted with a resistance. The detection of *n*=1-4 photons in the telecom wavelength range was demonstrated in a 4-elements PNR with maximum system quantum efficiency of 2.6% and recovery time in the ns range. Scaling to large number of wires and integration of cavity or waveguide structures should enable efficient PNRs for linear detection in the few to few tens of photons range.

This research is supported by EU-FP7 projects QUANTIP (project No. 244026) and Q-ESSENCE (project No. 248095), the Dutch Technology Foundation STW, applied science division of NWO, the Technology Program of the Ministry of Economic Affairs (project No. 10380), and NanoNextNL, a micro and nanotechnology program of the Dutch ministry of economic affairs, agriculture and innovation (EL&I) and 130 partners.